\documentclass[10pt,a4paper]{article}
\usepackage{booktabs}
\usepackage{amsmath}
\usepackage{amsfonts}
\usepackage{amssymb}
\usepackage{graphicx}
\usepackage{fancyhdr}
\usepackage{gensymb}
\usepackage{fixltx2e}
\usepackage{authblk}

	\title{Quantitative study  of the pinning effect of the  edge dislocation on  domain wall motion in Barium Titanate thin films}
	
	\author[a]{Shuai Wang}
	\author[a]{Li-Hua Shao  \thanks{Corresponding author:shaolihua@buaa.edu.cn}}
	\affil[a]{National Key Laboratory of Strength and Structural Integrity, Institute of Solid Mechanics, School of Aeronautic Science and Engineering, Beihang University, Beijing 100191, China }

\begin{document}
		\maketitle
		\begin{abstract}
			 Dislocation is a very important one-dimensional defect in ferroelectrics.
			This work introduces  an easy and flexible model
			of implementing the edge dislocation by introducing eigenstrain at the interface, and it could be easily extended to
			incorporate the surface stress to refine the analysis of ferroelectric thin films. The influence of dislocations on the ferroelectric domain wall motion and hysteresis loop including the remanent polarization and 	coercive field using phase-field simulations is analyzed. The pinning effect of the dislocation on the domain wall motion is discussed and whether the domain wall is pined is the competition between the external loading and the magnitude of the burgers vector of the dislocation. This work could contribute to the understanding of the pining effect of the dislocation  and provide guidance for the fabrication of ferroelectric thin films.
		\end{abstract}


Ferroelectrics are essential components in a wide spectrum of applications due to their unique electro-mechanical coupling properties.~\cite{scott2007applications} 
The nonlinear properties  of the electromechancial coupling effect of ferroelectrics originate from the spontaneous polarization.
The spontaneous polarization can be switched by external field or mechanical loading   via the nucleation and growth of a more energetically favored domains through a highly inhomogeneous process, whereby local variations in free energy from defects dominate the switching kinetics.~\cite{gao2011revealing}
The ferroelectric domain wall represents the transition region of the polarization  between neighbouring domains and the domain wall motion is responsible for the  non-linear dielectric, piezoelectric and elastic properties of ferroelectrics. Defects and material inhomogeneities such as oxygen vacancies, point defects and  dislocations  play a crucial role when explaining the abnormal phenomenons during the domain wall motion such as pining effect at the micro scale.  
The pining effect of the domain wall could be found in either bicrystal grain boundary~\cite{rodriguez2008ferroelectric} polycrystalline gain boundaries~\cite{schultheiss2020domain} or at the defect regions in thin films. For instance,  Dragan \textit{et al.} provide experimental evidence that the domain-wall movement in BiFeO\textsubscript{3} is strongly inhibited by charged defects and the domain-wall mobility can be considerably increased by preventing the defects from migrating into their stable configuration.~\cite{rojac2010strong}

Compared with their bulk counterparts, ferroelectric thin films are essential components in  applications at micro/nano scale,   such as microsystems~\cite{defay2013integration}, memory devices~\cite{mikolajick2020past},
and high frequency electrical components\cite{zou2022aluminum}. 
The divergence between the properties of bulk ferroelectrics and thin films comes from the surface effect, large depolarization field, film/substrate interface, epitaxial stress and so on.
Dislocations are one-dimensional defects that commonly exist
in polycrystallines  and thin films.~\cite{cai2018imperfections}
The dislocation densities in bulk ferroelectric single crystals~\cite{hofling2021large} and ceramics mainly affect the mechanical properties and hysteresis through the plastic deformation. 
While for the epitaxial  films and heterostructures, the dislocations are almost unavoidable and has a direct influence on the   domain topology.
The dislocations in thin films has been extensively studied.~\cite{nix1989mechanical}
The strain due to the misfit strain 
between the thin film layer and a substrate may be accommodated by the misfit dislocations at the interface.~\cite{matthews1974defects} The role of the misfit dislocation is to release the mechanical energy caused by the misfit strain. The existence of the dislocations make the ferroelectric thin films excellent
candidates for innovative device concepts, ranging from dislocation-based  memory devices to light emission diodes~\cite{lester1995high}. 
		
The role of the dislocations  on the domain wall motion and macroscopic properties in ferroelectric thin films has been extensively studied in the past decades. Dai \textit{et al.}~\cite{dai1996link} first point out the that the dislocations act as pinning sites for the domain boundaries. Antonios and Landis~\cite{kontsos2009computational} use a phase-field method to systematically investigate the influence of the edge dislocation on the domain configuration. In their model, the dislocation regions are represented by the mechanical/electric boundary conditions. In a recent series of works by Zhou et al.~\cite{zhou2022influence,zhou2022phase,zhuo2023intrinsic} regard dislocations as order parameter to study the interaction between dislocation and domains. Cheng et al. combines STEM and 2-dimensional phase-field simulation and find dislocation pairs are more favorable for the retention of a domains by playing a pinning role.~\cite{cheng2023dislocations} Jiang et al. observes the pining effect of the dislocation on the domain wall in BTO thin films.~\cite{jiang2023observation}

In the work by Kröner~\cite{kroner1981continuum},   the elastic field with defects in solids is  mathematically
described, which led to solving the stress field by taking the dislocation induced strain as eigenstrain.
In this letter, we perform a quantitative analysis of the pinning effect of the  edge dislocation on  domain wall motion in ferroelectric thin films. Different from the work of Landis or Zhou, the   dislocation in thin film is represented by the misfit eigenstrain at the film-substrate interface. The influence of the dislocation on the hysteresis loops and the domain wall motion is quantitatively analyzed. Finally the criteria of the domain wall-dislocation pinning is given to give a guidance for the ferroelectric thin film fabrication and dislocation-engineered domain tuning.


To model the influence effect of the dislocation on the ferroelectric thin film, three assumptions are proposed:\\
(1) Only the edge dislocation occurs in the ferroelectric thin film;\\
(2) The position of the dislocation do not change during the domain evolution;\\
(3) The influence of the dislocation on the polarization mainly originates from the stress field of the dislocation.\\
We assume a right-handed Cartesian coordinate system centered at the core of a single, straight dislocation. The dislocation is characterized 
by the dislocation line vector $\xi$ (with unit length) and the Burgers vector $\boldsymbol{b}$ (with length b ). For simplicity, we consider in the current paper only edge dislocations with a line vector oriented in 
z-direction, $\xi=  [0, 0, 1] $, and a Burgers vector pointing into the x-direction, $\boldsymbol{b}=  [0, 0, b] $. The dislocation eigenstrain  $\varepsilon_{ij}^{\text{dis}}$  is prescribed as one row finite element mesh  on the slip plane in the x-direction. The normal component of $\varepsilon_{ij}^{\text{dis}}$ is zero and the shear component of  $\varepsilon_{ij}^{\text{dis}} =\frac{b}{l_0}$, where $l_0$ is the width of the mesh in the y-direction (see Fig. 1(a)). Here we use a uniform size of a four node  linear Lagrange element throughout the paper.

Figure 1 shows the contour plots of the stress components $\sigma_{xx}$, $\sigma_{yy}$ 
and $\sigma_{xy}$ around the single edge dislocation in the unloaded body. The  analytical solutions are taken from the literature~\cite{lubarda2007configurational}.
It is seen that the dislocation model proposed here produces stresses that coincide very well with the analytical solution. The normal stress components  coincide equally well.  The dislocation core is the singularity of the stress field. Differences between 
modeling results and analytical solution only occur near the dislocation core.

\begin{figure}[!hbp]
	\centering
	\includegraphics[width=\textwidth]{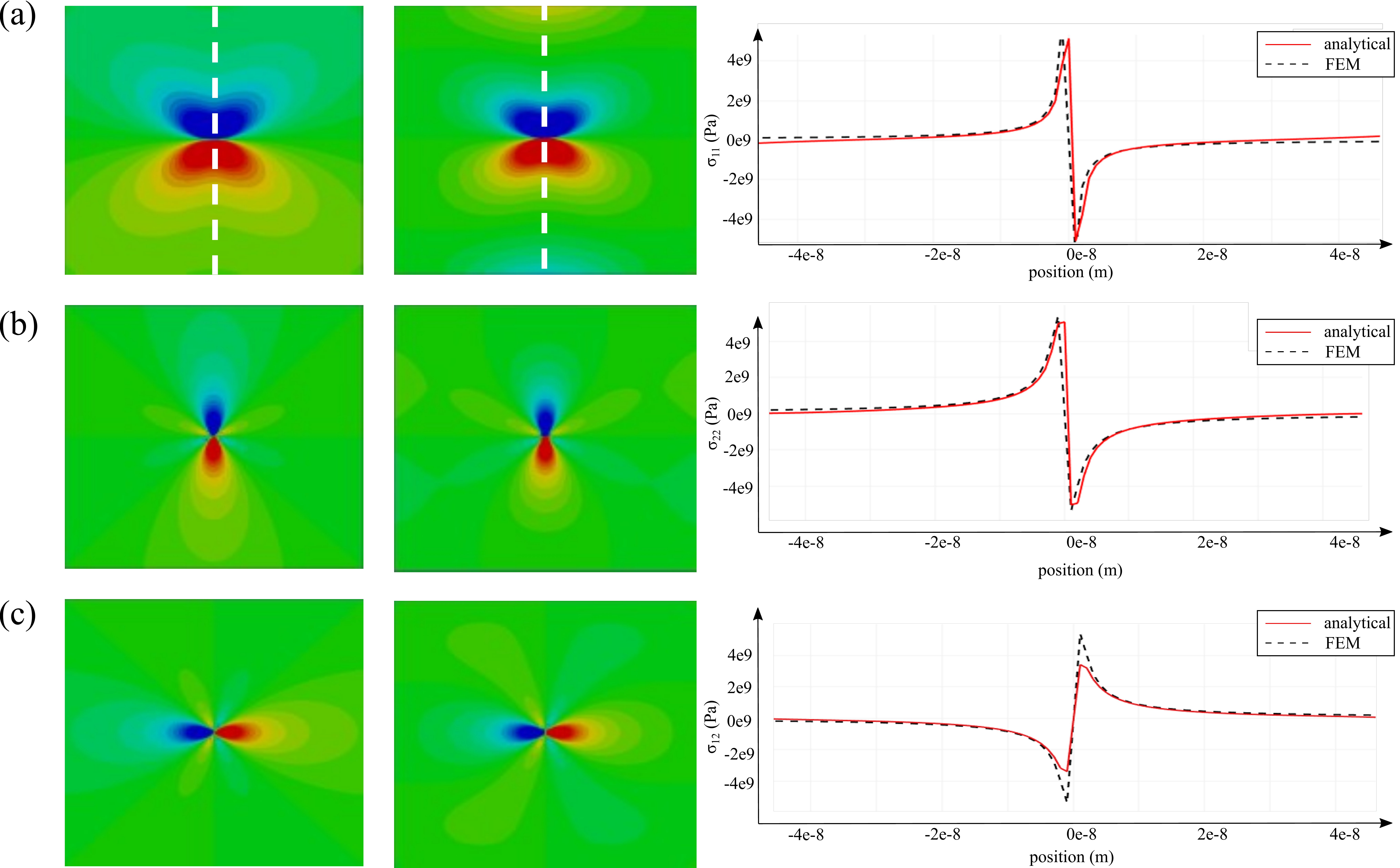}
	\caption{Contour plots of the stress components  around a single edge dislocation in an unloaded body: Comparison between analytic solutions:
		(a) $\sigma_{xx}$
		(b) $\sigma_{yy}$
		(c) $\sigma_{xy}$.}
	\label{fig1}
\end{figure}

Phase-field models have been used extensively to describe various aspects of the  behaviors of the ferroelectric thin films.~\cite{chen2008phase} 
The framework of the ferroelectric model in this letter is based on the previous work of Xu \textit{et al.}~\cite{xu_phase_2010} and Wang \textit{et al.}~\cite{wang2016phase}. 	
The paraelectric-to-ferroelectric phase transition occurs in a ferroelectric material when its temperature is
lower than its Curie point. The  spontaneous polarization $\boldsymbol{P}$ is adopted as the order parameter.  In phase
field simulations, the time-dependent Ginzburg–Landau
equation is used to describe the polarization evolution and thus the domain configuration change,
\begin{equation}
	\frac{\partial P_i (\boldsymbol{x},t)}{\partial t}
	=-M\frac{\delta \mathcal H}{\delta P_i(\boldsymbol{x},t)} 
\end{equation}
where $M$ governs the mobility of the polarization vector, $\mathcal H$ is the electric enthalpy of the system.
	In the model, the electrical enthalpy is constituted with four parts, i.e.
\begin{equation}\label{flexoenergy}
	\mathcal{H}=\mathcal{H}^{ela}+\mathcal{H}^{ele}+\mathcal{H}^{coup}
	+\mathcal{H}^{bulk}+\mathcal{H}^{grad}
\end{equation}
in which $\mathcal{H}^{ela}$, $\mathcal{H}^{ele}$, $\mathcal{H}^{coup}$, 
$\mathcal{H}^{bulk}$ and $\mathcal{H}^{grad}$ represent elastic energy density, 
electrical energy density, mechanical-electric coupling energy density, 
Landau free energy density and gradient energy,	respectively. These energy densities are given in the following form,
\begin{equation}
	\left\{
	\begin{aligned}
		\mathcal{H}^{ela}&=\frac{1}{2} c_{ijkl} \varepsilon_{ij}^{e} 
		\varepsilon_{kl}^{e}\\
		\mathcal{H}^{ele}&=-\frac{1}{2} k_{ij} E_i E_j-P_iE_i\\
		\mathcal{H}^{coup}&=-b_{ijk} \varepsilon_{ij} E_k\\
		\mathcal{H}^{bulk}&=\beta_1(G,\lambda)(a_{ij}P_iP_j+a_{ijkl}P_iP_jP_kP_l)+a_{ijklmn}P_iP_jP_kP_lP_mP_n\\
		\mathcal{H}^{grad}&=\beta_2(G,\lambda)( P_{i,j} P_{k,l})	
	\end{aligned}
	\right.
\end{equation}
where 
$\varepsilon_{ij}^{e}=(\varepsilon_{ij}
-\varepsilon^0_{ij}(\boldsymbol P)
-\varepsilon^\text{dis}_{ij}
)
$ 
is the elastic contribution to the  strain. 
$\varepsilon^{i}_{ij}$ is the total strain and 
$\varepsilon^0_{ij}(\boldsymbol 
P)$ is the remnant strain induced
by the remnant polarization, $\beta_{1}$ and $\beta_{2}$ are the coefficients 
related to the domain wall energy $G$ and the domain wall thickness 
$\lambda$\cite{schrade2008phase}. Here 
$\varepsilon^0_{ij}(\boldsymbol P)$ are calculated following 
the 
work by Huo and Jiang \cite{huo1997modeling}.
\begin{equation}
	\varepsilon^0_{ij}(\boldsymbol P)=\frac{3}{2}\varepsilon_{sat} 
	\frac{\sqrt{P_iP_i}}{P_{sat}}(n_in_j-\frac{1}{3}\delta_{i,j})
\end{equation}
where $n_i$ is the unit vector of $P$, $\varepsilon_{sat}$ is the maximum 
remnant strain and $P_{sat}$ is the maximum remnant polarization.
Strain tensor $c_{ijkl}$ and permittivity tensor $k_{ij}$ are the same 
as macroscopic ones. For piezoelectric tensor $b_{ijk}$, we used the 
representation 
\begin{equation}
	\begin{split}
		b_{ijk}(\boldsymbol P)=&  \frac{\sqrt{P_iP_i}}{P_{sat}} \{
		d_{33}n_in_jn_k+d_{31}(\delta_{ij}-n_in_j)n_k \\
		&+\frac{1}{2} d_{15}\left[  (\delta_{ki}-n_kn_i)n_j+(\delta_{kj}-n_kn_j)n_i))
		\right]
		\}
	\end{split}
\end{equation}
adopted by Kamlah\cite{kamlah2001ferroelectric}.

\begin{figure}[!hbp]
	\centering
	\includegraphics[width=0.8\textwidth]{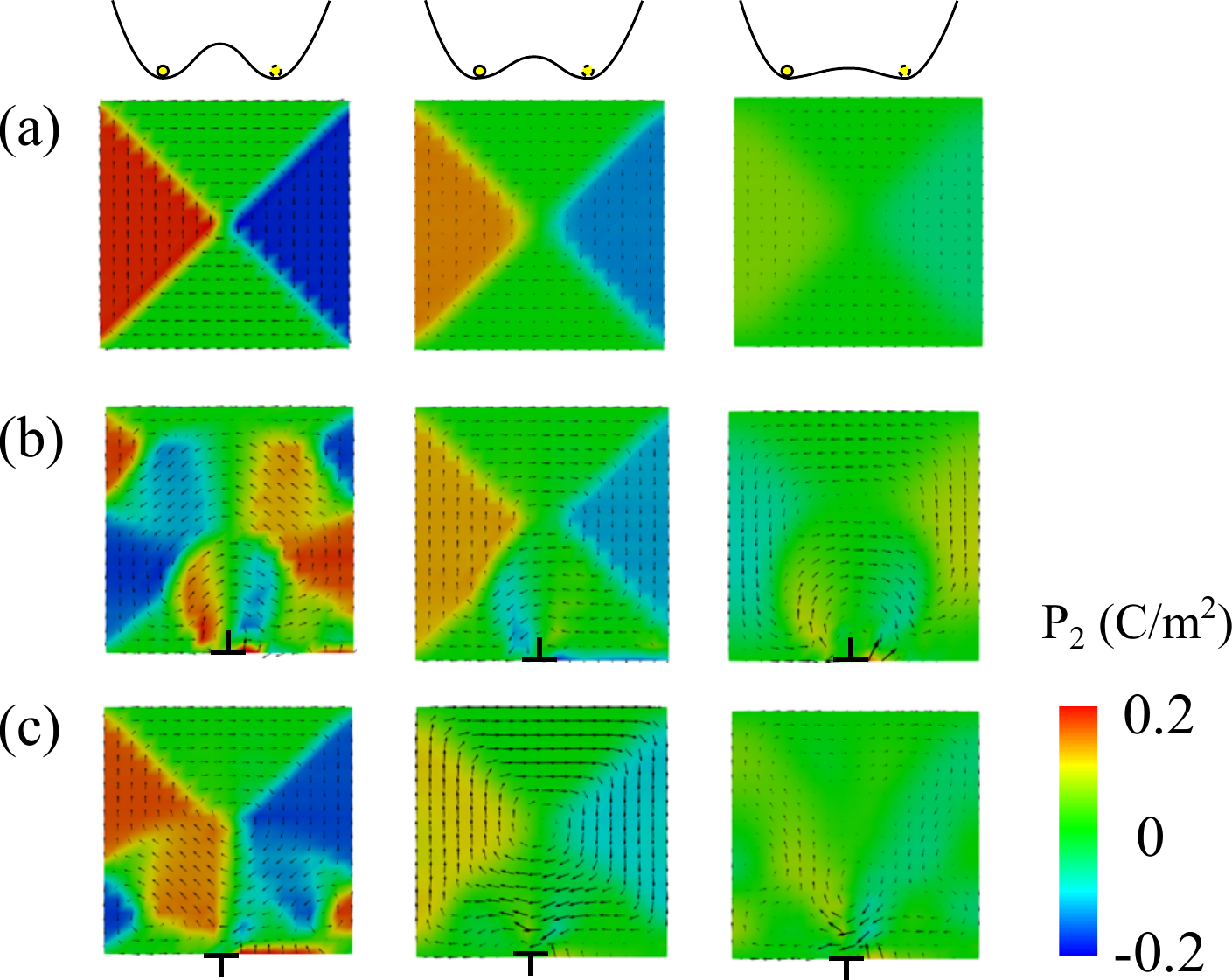}
	\caption{ Influence of the edge dislocation on the domain vortex at different temperature.
		(a)without the edge dislocation.
		(b) with the edge dislocation at the bottom boundary of the thin film. (Burgers vector pointed to the right)
		(c) with the edge dislocation at the bottom boundary of the thin film. (Burgers vector pointed to the left)
		from left to right: 300 K, 500 K, 800 K.
	}
	\label{fig2}
\end{figure}

 Here we use the parameters for barium titanate for the simulation.~\cite{chen2007appendix}
In the 
Figure 2 shows the domain configuration of a BTO thin film with different edge dislocation at  temperature ranges from 300 K 500 K to 800 K. The change-free boundary and stress-free condition is set for all boundaries and the initial polarization distribution is set random. Figure 2(a) shows that the final configuration of the domain is a vortex. As the temperature rises,  the magnitude of the polarization decreases and  the energy barrier is lowered, which indicates that the domain configuration is very sensitive to the external stimuli at elevated temperature. If the edge dislocation is introduced to the system (see Fig. 2(b) and Fig. 2(c)), the vortex is twisted near the core and the dislocation with burger's vector pointed to the right pushes the polarization to the vortex center, while the dislocation in with the burger's vector in the opposite direction attracts the polarization near the dislocation core. As the temperature rises, the vortex become less distinct.

\begin{figure}[!hbp]
	\centering
	\includegraphics[width=\textwidth]{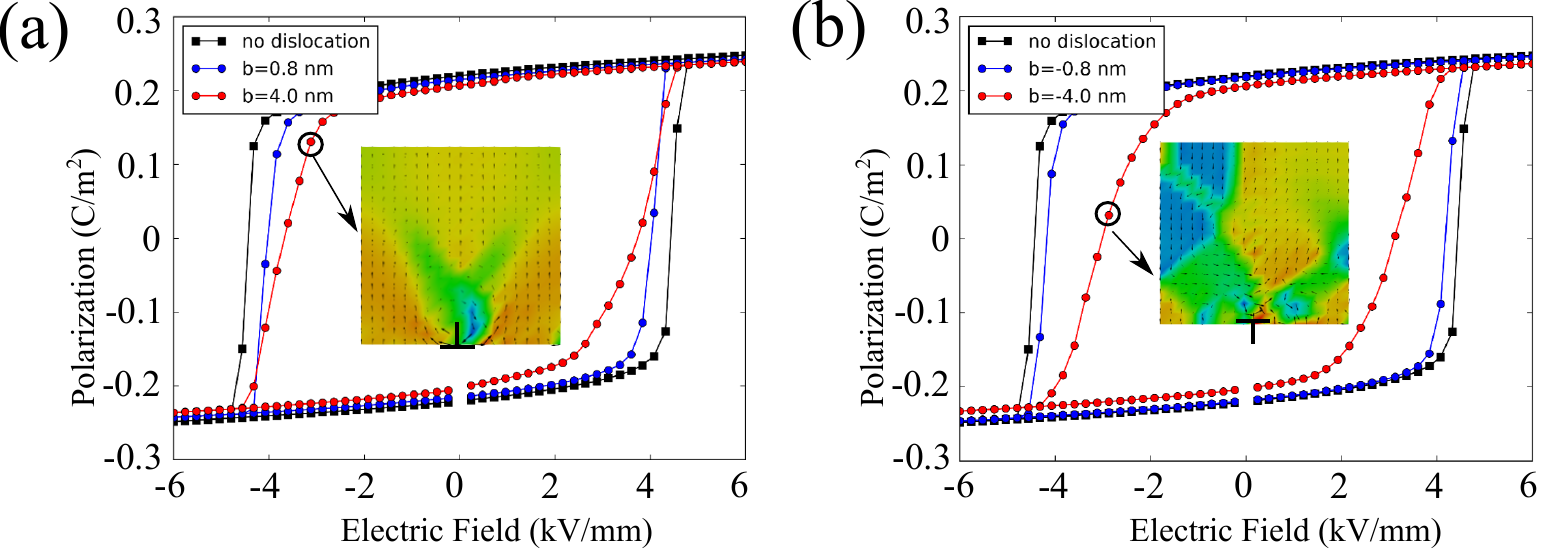}
	\caption{Influence of edge dislocation on the polarization hysteresis loop of the BTO thin film. (a) Burgers vector pointed to the right. (b) Burgers vector pointed to the left }
	\label{fig3}
\end{figure}

Figure 3 shows influences of the dislocation with different direction on the  the hysteresis loop. The hysteresis becomes slimmer as the magnitude of the burgers vector increases regardless the direction of the burgers vector.  The domain configuration at the same electric field (3 KV/mm) is plotted as the inset figures. Compared with Fig. 3(a), the domain configuration in Fig. 3(b) is more randomized due to the attraction effect from the dislocation core, which  makes the 
overall polarization smaller.

\begin{figure}[!hbp]
			\centering
			\includegraphics[width=0.5\textwidth]{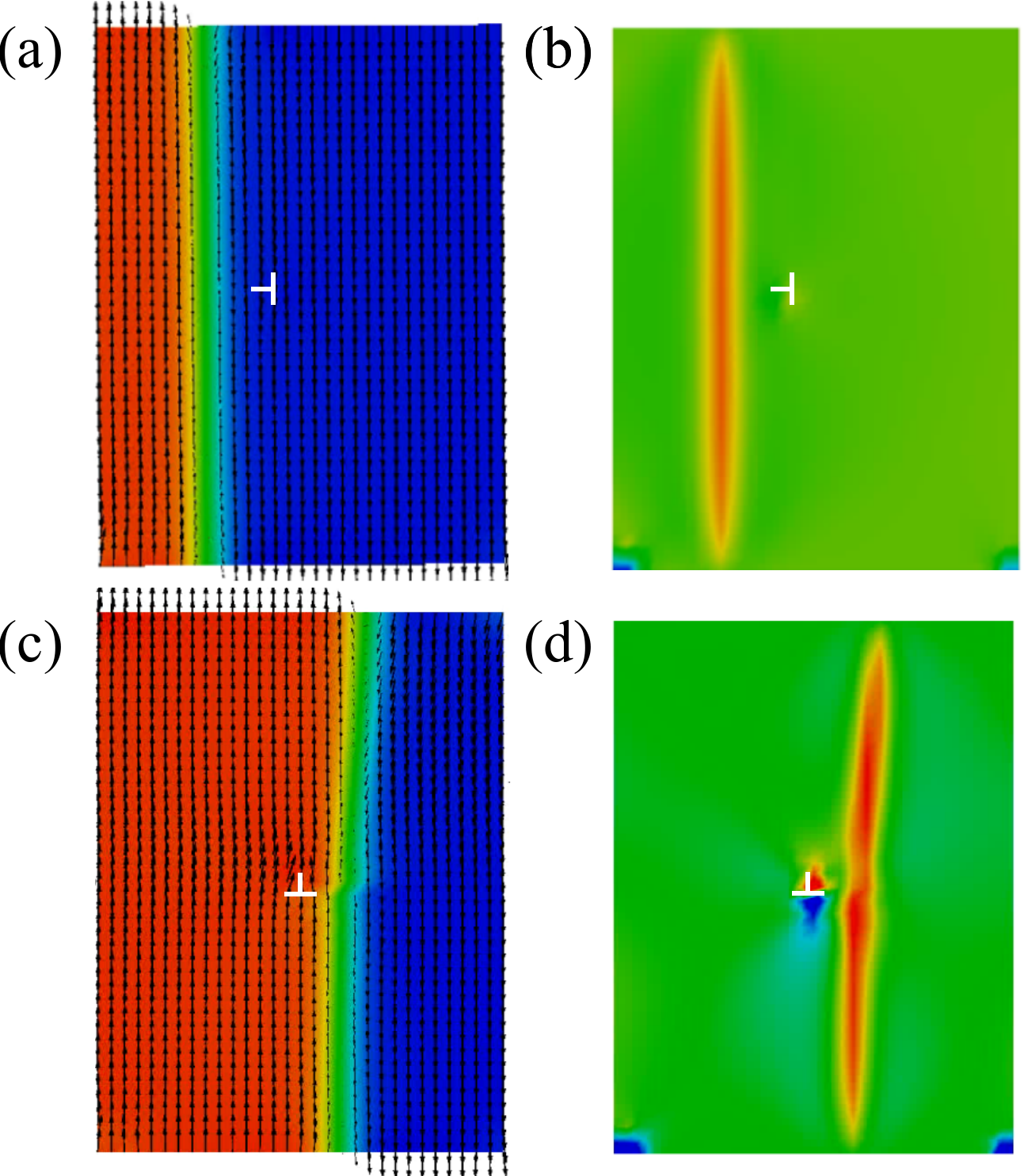}
			\caption{Domain wall motion at the edge dislocation with direction. (a)Burgers vector parallel to the domain wall, legend: $P_y$ (b) Burgers vector perpendicular to the domain wall, legend: $\sigma_{yy}$ }
			\label{fig4}
\end{figure}

The last simulation shows the domain wall motion at the edge dislocation. At the initial state, a 180\degree	domain wall is assumed in the y-direction which separates the left domain (upright) from the right domain (downward). An electric potential is given at the top  surface to give the driving force for the domain wall motion. Here we use the coercive field $E_c$ as the  reference field value and one unit cell length ($b_0= a = 0.4$ nm) as the reference burgers vector's magnitude.

We first examine the effect of the edge dislocation with burgers vector pointed to the positive y-direction on the domain wall motion.
 For the applied electric field  $E = 3 E_c$ and $b = b_0$,  the domain wall motion is hindered at the dislocation core.
 While for the case where the burgers vector pointed to the positive x-direction, the domain wall will pass through the dislocation core. This proves that not only the magnitude but also the direction of the edge dislocation determines the difficulty of domain switching. For the first case shown in Fig.4(a), if one increase the external field to $E = 6 E_c$. The domain wall can pass through the edge dislocation call. More simulation cases can be found in the Supplementary Material. To further investigate the antagonism
 between the strength of the dislocation and the external field strength, a series of simulation is carried out and whether the domain wall successfully pass through the dislocation core is illustrated in Fig. 5. The burgers vector and the external fields are all normalized by $b_0$ and $E_c$ ($b^\star =b/b_0$ and $E^\star=E/E_c$). For the case without external field, the domain wall does not move.
 At the coercive field ($E^\star=1$), the domain wall can only move without edge dislocation.
  For the same burgers vector, there is a threshold field where the domain wall can pass the dislocation core. For the same external field stimuli, the domain wall can only pass the dislocation core at the lower burgers vector. 
 We use the nonlinear regression algorithm  fit the watershed as a dash line in Fig. 5. The exponential fitting of the watershed is $b^\star = -0.088+(0.024E^\star)^{1.12}$. The slope of the curve is much smaller when $E^\star < 3$.
 
 \begin{figure}[!hbp]
 	\centering
 	\includegraphics[width=0.5\textwidth]{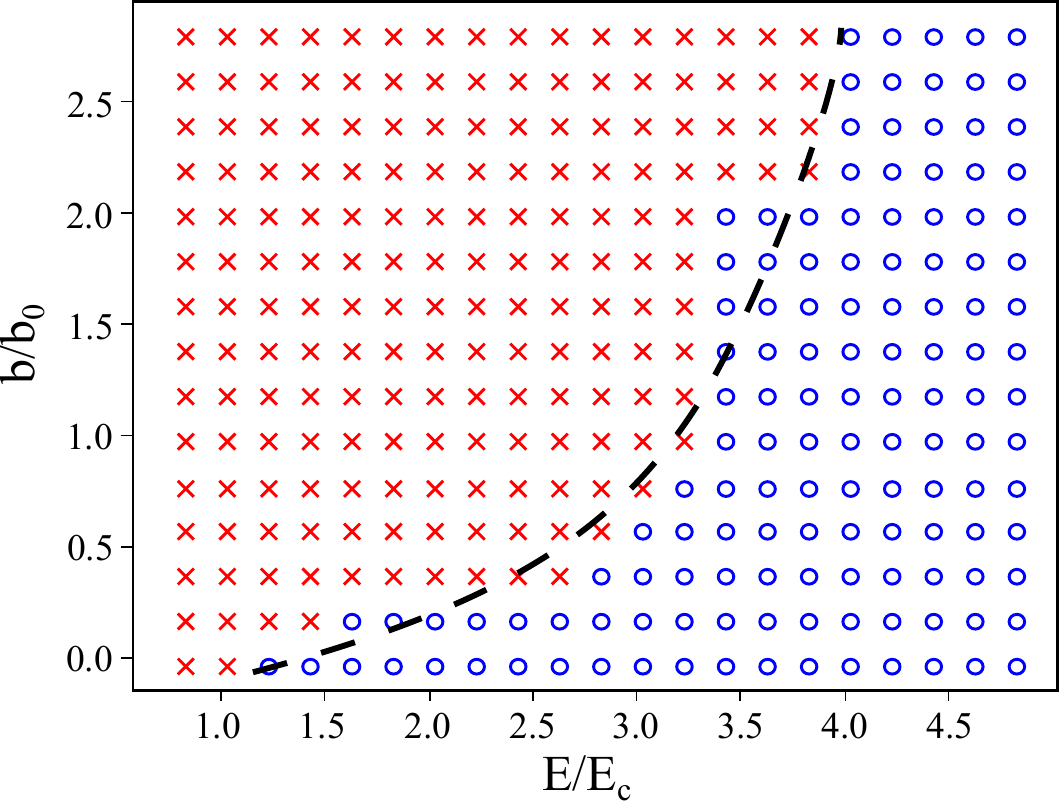}
 	\caption{
 		Threshold of Burgers vector strength for  pinning effect of the dislocation on the domain wall motion. The red cross mark represent the simulation case that the domain wall does not pass through the edge dislocation and the hollow circle indicates that the  domain wall pass through the edge dislocation.
 	}
 	\label{fig5}
 \end{figure}

In summary, this study provides a simple while effective way to implement the edge dislocation to the phase field ferroelectric thin film model. The competitive of the external electric field loading and edge dislocation burgers vector is quantitative studied and the criteria for the domain wall pinning is given. Normally for the epitaxial growth ferroelectric thin film, the burgers vector is one unit cell and the distance between two dislocation cores are far enough that one can ignore the interaction between two different dislocations. The proposed criteria  may guide the domain wall tuning for the thin  films where the dislocations is introduced. The above exponential relation between $E^\star$ and $b^\star$ is only valid for barium titanate. For other material systems, the parameters should be tuned since the energy barrier high of the polarization switching and the piezoelectric coefficients are different. Moreover, since the stress/strain field near the dislocation core changes dramatically and the strain-gradient effect and flexoelectric effect should be considered for a more accurate analysis.

	\section*{Conflict of Interest}

	The authors have no conflicts to disclose.	
	
\section*{Data availibility}
	
	The data that support the findings of this study are available within the article and its supplementary material.
		
		\section*{Acknowledgments}
		This work was supported by the Beijing Natural Science Foundation (no. JQ21001) and National Natural Science Foundation of China (NSFC no.12272020). S.W. acknowledges the support by “the Fundamental Research Funds for the Central Universities” (YWF-23-SDHK-L-019). The authors acknowledge the High Performance Computing Platform of Beihang University for partial support of this work.

		\bibliographystyle{unsrt}
		\bibliography{ref.bib}
	\end{document}